# AC losses in Bi,Pb(2223) multifilamentary wires with square cross-section

G. Witz[a], X.-D Su[a], K. Kwasnitza[c], and R. Flükiger[a,b]

[a] GAP, University of Geneva, 20 rue de l'Ecole de Médecine, CH-1211 Genève, Switzerland
[b] DPMC, University of Geneva, 24 quai Ernest Ansermet, CH-1211 Genève, Switzerland
[c] CH-8173, Riedt, Switzerland

**Abstract:** We have fabricated prototype Bi,Pb(2223)/Ag superconducting multifilamentary wires with a square cross-section. The AC losses of these wires were measured, compared with those of tapes and also compared with the theory. Wires show largely reduced AC magnetic field losses at 47 Hz, compared to those of tapes in a perpendicular magnetic field. The effects of the twist pitch and of the use of the high resistive barriers were also investigated. The results show that the use of a twist pitch of about 10 mm is sufficient for decoupling the filaments, and that the use of very short twist pitches does not further reduce the absolute value of the AC losses. The losses in the wires are still higher than those of tapes in a parallel magnetic field, but the results show that the square or round configurations are interesting for applications where a perpendicular component of the magnetic field cannot be avoided, or for applications in rotating magnetic fields.

## 1 Introduction

In this introduction we will present the state of the art in the field of High $T_c$ superconductors for AC applications and some theoretical aspects of the magnetization losses due to the coupling currents. Twisted Bi,Pb(2223)/Ag multifilamentary tapes with a typical aspect ratio of 15 have relatively large coupling current magnetization losses at 50 Hz in a perpendicular applied magnetic field. This is related also to the large inductance due to the large wide side $c$ of the tape and to their small resistance given by the small thickness of the tape, this is the direction in which the coupling currents flow through the normal-conducting matrix.

For the case where filament saturation effects by the induced coupling currents are not yet present, one has the following equation for the coupling current power loss per volume unit in a sinusoidal magnetic field change [1]:

$$P = \frac{1}{29}\frac{c^2}{d^2}\frac{(c+3d)^2}{(c+d)^2}\frac{l_t^2}{\rho_e}\Delta B^2 \nu^2$$

$$P = \frac{1}{29}k\Delta B^2 \nu^2 \qquad (1)$$

where $c$ is the tape width, $d$ is the tape thickness, $l_t$ is the twist pitch, $\rho_e$ is the effective transverse electrical resistivity between the filaments, $\Delta B$ is the peak-to-peak magnetic field amplitude applied in perpendicular to c and $\nu$ is the frequency.
For d << c this gives:

$$P = \frac{1}{29}\frac{c^2}{d^2}\frac{l_t^2}{\rho_e}\Delta B^2 \nu^2 \qquad (2)$$

There is a considerable interest to use Bi,Pb(2223)/Ag tapes for applications like superconducting transformers at 50 Hz and at typical $\Delta B = 0.2$ T. In such magnet applications, both magnetic field losses and the transport current self-field losses are present. But the magnetic losses are dominant [2] as long as the transport current is not extremely close to $I_c$.

It has been found experimentally that in flat tapes with a pure Ag matrix and a twist pitch of about 1-2 cm the coupling losses are more than one order of magnitude larger than the tolerable loss level [3] of $P \approx 0.45$ mW/Am at 50 Hz and $\Delta B = 0.2$ T. At 50 Hz full composite saturation by the induced currents occurs and the whole composite behaves like a single filament with large losses of the hysteretic type. This was the reason for the development of twisted Bi,Pb(2223) multifilamentary tapes having highly resistive ceramic barriers of $BaZrO_3$ or $SrZrO_3$ between the filaments in the Ag matrix [4]-[7]. We found that the barriers increase the value of $\rho_e$ by about a factor of 10 – 15 compared to the resistivity of Ag ($\rho_{Ag}$), while the coupling current decay time constant $\tau$ could be reduced to about 0.9 ms using a twist length of 13 mm. The losses for $\Delta B$ applied parallel to the tape width could be reduced to values below the tolerable loss level mentioned above, which motivated a series of works on Bi,Pb(2223) tapes with various oxide barriers [8]-[10]. An Ag-Au matrix was used [10], also together with $SrZrO_3$ barriers [8]. In combination with $l_t$ values smaller than 10 mm, $\tau$ values down to 0.3 ms could be realized. However, this development direction did not succeed in reaching sufficiently low losses for $\Delta B$ perpendicular to the tape surface and 50 Hz. The reasons are the following ones:
- The typically 2 $\mu$m thick ceramic barriers have holes and interruptions
- The Ag-10%Au matrix increases the $\rho_e$ only by a factor of about 8. In addition, Ag-Au is too expensive for tape fabrication

In conclusion one can say that in barrier tapes with $l_t \geq 8$ mm, there is already a considerable loss reduction at 50 Hz for intermediate $\Delta B < 0.1$ T values. But for $\Delta B \geq 0.1$ T the losses are clearly above the mentioned loss level and composite saturation occurs. In this saturation case, one has for the losses per cycle and volume unit for

$\Delta B \gg \Delta B_{ps}$, where $\Delta B_{ps}$ = saturation field of the composite:

$$Q_{sat} = \frac{1}{2} \Delta B \cdot \overline{J_c} \cdot c \qquad (3)$$

$\overline{J_c}$ is the value of the critical current density averaged over the cross section of the filamentary zone. In a square conductor with the same cross-section area as that of a flat tape, the width c is much smaller. Even in the case of full composite saturation, according to eq. (3), the square conductor will have several time smaller saturation losses than the flat tape.

Concerning the present situation for the barrier tapes, it turns out that in order to get sufficiently low AC loss conductors, one should still reduce the product

$$k = \frac{c^2}{d^2} \frac{(c+3d)^2}{(c+d)^2} \frac{l_t^2}{\rho_e} \qquad (4)$$

roughly by a factor of 50-80. This goal can be partly achieved by replacing the rectangular flat tape by wires with square cross sections. We have thus started the development of Bi,Pb(2223)/Ag multifilamentary conductors with square overall cross sections, having an Ag matrix. The case of additional ceramic barriers was also studied. Until now, prototypes of square and round wires on the basis of Bi(2212) have been developed in several laboratories, and AC loss measurements were also performed on multifilamentary Bi,Pb(2223) round wires with small superconductor filling factors [11]. There is a criterion where composite saturation becomes significant in twisted multifilamentary tapes. For flat tape with $d \ll c$, assuming for simplicity $l_t \gg c$, saturation of the outer filament layer occurs [6] at a critical rate of magnetic field change

$$\dot{B}_c = 48 \cdot J_c \cdot \rho_e \cdot b \cdot \frac{d}{c} \frac{1}{l_t^2} \qquad (5)$$

where b is the filament thickness. The general equation for any c and d values is

$$\dot{B}_c = 48 \cdot J_c \cdot \rho_e \cdot b \cdot \frac{d}{c} \frac{(c+d)}{(c+3d)} \frac{1}{l_t^2} \qquad (6)$$

Setting for the square conductor $c = d$, eq. (6) becomes

$$\dot{B}_c = 24 \cdot J_c \cdot b \frac{\rho_e}{l_t^2} \qquad (7)$$

Here $\dot{B}_c$ is much larger and therefore more favourable than the corresponding value for a flat tape. For instance, with $c/d = 15$, eq. (6) gives only the 7.5 times smaller values

$$\dot{B}_c = \frac{16}{5} \cdot J_c \cdot b \frac{\rho_e}{l_t^2} \qquad (8)$$

For a square conductor with $c = d$, the coupling loss in eq. (1) becomes

$$P = \frac{1}{7.3} \frac{l_t^2}{\rho_e} \Delta B^2 v^2 \qquad (9)$$

This is close to the coupling loss equation for a round wire [12], where

$$P = \frac{2}{\mu_0} \dot{B}^2 \cdot \tau \qquad (10)$$

with

$$\tau = \frac{1}{2} \frac{1}{4\pi^2} \mu_0 \frac{l_t^2}{\rho_e} \qquad (11)$$

Averaging over a sinusoidal field change, one gets

$$\overline{\dot{B}_c^2} = \frac{1}{8} \Delta B^2 \cdot 4\pi^2 \cdot v^2 \qquad (12)$$

which leads to

$$P = \frac{1}{8} \frac{l_t^2}{\rho_e} \Delta B^2 v^2 \qquad (13)$$

Comparing the equations (9) and (13) shows that the square conductor with side length c has only about 10% larger losses than the round wire with diameter c. Comparing further the coupling losses of flat twisted tapes with an aspect ratio of about $c/d = 15$ to those of a square wire with the same $l_t$ and $\rho_e$ we expect from the equations (1) and (9) would give about 70 times smaller coupling losses for the square conductor in the case of non-saturation.

It is also interesting to compare the $\tau$ values in both geometries. For the tape of rectangular cross section, one has for $\tau$ [1]:

$$\tau = \frac{1}{144} \frac{c^2}{d} \frac{(c+3d)^2}{(c+d)^3} \mu_0 \frac{l_t^2}{\rho_e} \qquad (14)$$

This gives for a very flat tape with $d \ll c$

$$\tau = \frac{1}{144} \frac{c}{d} \mu_0 \frac{l_t^2}{\rho_e} \qquad (15)$$

For a wire with a square cross section, equation (14) reduces to

$$\tau = \frac{1}{72} \mu_0 \frac{l_t^2}{\rho_e} \qquad (16)$$

This again is very close to the $\tau$ expression for the round wire in equation (11)

$$\tau = \frac{1}{2} \frac{1}{4\pi^2} \mu_0 \frac{l_t^2}{\rho_e} \approx \frac{1}{79} \mu_0 \frac{l_t^2}{\rho_e} \qquad (17)$$

It follows that the development of multifilamentary Bi,Pb(2223)/Ag conductors with square cross section (eventually also with barriers) will lead very close to the required low AC loss conductor, at least for transformer applications.

Table I
Wire characteristics

| Wire | Barrier | No of filaments | Wire size (mm) | Filamentary zone size (mm) | Sample length (mm) | Twist pitch (mm) | $I_c$ (A) | $J_c$ (A/cm$^2$) | Measurment configuration |
|---|---|---|---|---|---|---|---|---|---|
| A | - | 40 | 1.1x1 | 0.9x0.8 | 20 | $\infty$ | 8 | 3600 | 3x3 |
| A | - | 40 | 1.1x1 | 0.9x0.8 | 8 | 8 | 7.2 | 3300 | 3x3 |
| A | - | 40 | 1.1x1 | 0.9x0.8 | 8 | 4 | 6.3 | 2800 | 3x3 |
| A | SrZrO$_3$ | 36 | 1.1x1 | 0.8x0.75 | 8 | 8 | 3.2 | 2200 | 3x3 |
| B | - | 36 | 1x0.95 | 0.85x0.8 | 12 | $\infty$ | 13.5 | 7800 | 3x3 |
| B | - | 36 | 1x0.95 | 0.85x0.8 | 12 | 4 | 8 | 4600 | 2x2 |
| B | - | 36 | 1x0.95 | 0.85x0.8 | 12 | 2 | 5.5 | 3200 | 2x2 |
| C | - | 48 | 0.95x0.85 | 0.75x0.55 | 20 | $\infty$ | 11 | 11200 | 3x3 |

## 2 Experimental details

Square Bi,Pb(2223) wires, called A, with 40 filaments without barriers were prepared by restacking monofilamentary tapes in a square tube, using the configuration shown in Fig. 1. The tube was then two-axis rolled, drawn into a round shape with a diameter of 1.54 mm, twisted and then groove rolled down to a size of ≈1 mm of side. The wire was then heat treated, followed by an intermediate groove rolling deformation, and by the final heat treatment. A Bi,Pb(2223) wire with 36 filaments and barriers was prepared by restacking monofilamentary tapes covered with SrZrO$_3$ in a square tube, with the configuration shown in Fig. 1d, and then deformed and heat treated as the other wires. The barriers in this wire are still rather imperfect.

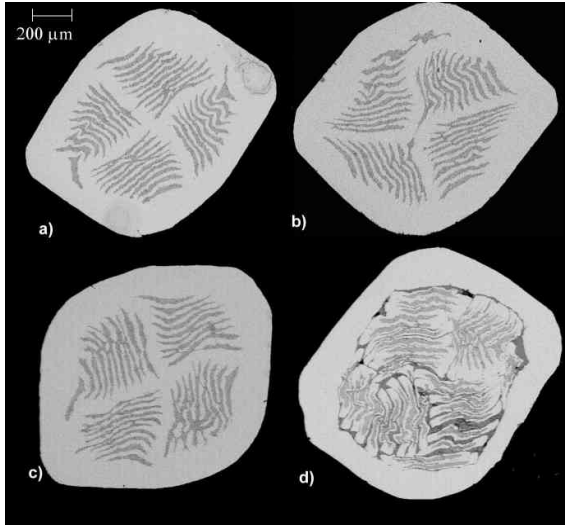

Fig.1 Cross-sections of wire A, a) untwisted without barrier, b) $l_t$ = 8 mm without barrier, c) $l_t$ = 4 mm without barrier, d) $l_t$ = 8mm with SrZrO$_3$ barriers.

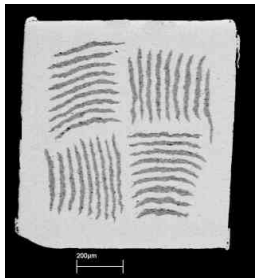

Fig. 2 Cross-section of untwisted wire B.

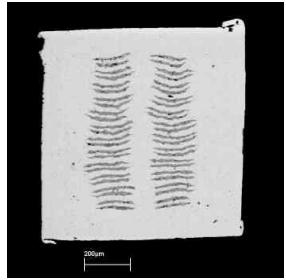

Fig. 3 Cross-section of wire C.

A second set of wires [13], called B, was prepared by restacking 36 monofilamentary tapes in a square tube, using the configuration shown in Fig. 2. After two-axis rolling down to a size of about 1 mm of side, the wires were heat treated, followed by an intermediate two-axis rolling deformation, twisting and a final heat treatment.

A third type of wire [13], called C, was prepared by restacking 4 monofilametary tapes into a square tube, which was then two-axis rolled. Then 12 pieces of the tube were restacked into another silver tube and two-axis rolled down to a size of about 1 mm. The wire was then heat treated, followed by an intermediate two-axis rolling deformation and a final heat treatment. In this conductor, the filaments are all parallel to each other, as shown in Fig. 3.

The characteristics of all the wires are compiled in the table 1. The AC losses in a perpendicular applied magnetic field change, were measured by a double Hall sensor method [7]. All measurements were performed at 77 K. The losses were measured on square stacks of 3x3 or 2x2 wires, which preserves the aspect ratio of the wire, the wires being isolated from each other in the stack. The lengths of the samples used for AC loss measurements were an integer multiple of the twist pitch and are listed in table 1. For the presentation of the losses per meter of conductor length, it was assumed that the losses occur mainly in the filamentary zone. In the figures, we display either the power losses per meter of conductor length as a function of $\Delta B$ or the power loss per meter of conductor length and ampere of critical current. For comparison a typical loss curve of an untwisted, 20 mm long sample of a Bi,Pb(2223)/Ag multifilamentary flat tape is shown (for perpendicular $\Delta B$). The tape (conductor C in [14]) has an aspect ratio of about 16 and a width $c$ = 3.6 mm.

The presentation of the power losses per meter of conductor length and ampere of critical current is more useful for the discussion on the applications, as the critical current varies in the differently twisted samples and also because of the engineering considerations. On the other side, the total magnetic field losses are the sum of the filament hysteresis losses, coupling losses and usual eddy current losses in the Ag. As long as the coupling currents do not saturate the conductor, the coupling losses are not proportional to $I_c$. So both presentations are useful.

## 3 Results and discussion

### 3.1 Square conductor A

#### 3.1.1 Magnetic field losses at 47 Hz

The square conductors A in Fig. 4 have 6 times smaller losses than the flat tape in a perpendicular magnetic field. Among the square conductor samples, the untwisted one with length $l$ = 20 mm exhibits the highest losses at smaller $\Delta B$ values. At the upper end its losses are proportional to $\Delta B^n$, with $n \cong 1.2$, while at the lower end one finds approximately $n \cong 2.9$. These $n$ values indicate that this sample is saturated at 47 Hz or is very close to saturation with coupling currents. Also the $n$ values of the flat tape with $n \cong 1.1$ at the upper end and $n \cong 2.9$ at the lower end indicate composite saturation losses. For the twisted samples with $l_t$ = 8 mm and 4 mm and for the sample with the still rather imperfect ceramic barriers and $l_t$ = 8mm, n values of 1.5, 1.9 and 1.8 were found at the highest $\Delta B$ range. This indicates that the induced coupling currents are much weaker and can not yet create composite saturation at 47 Hz in this $\Delta B$ range.

The measured loss results are also very interesting concerning their absolute values: the dashed horizontal line in Fig. 5 at 0.42 mW/Am (at 47 Hz) is the loss level limit $(P/I_c)_{tol}$ chosen for prototype High-$T_c$ transformer applications [3]. As Fig. 5 shows, the loss values of the square sample with $l_t$ = 4 mm are under this level for $\Delta B \leq 22$ mT, while for the flat tape, the corresponding threshold value is already about 5 mT. For transformer applications at 50 Hz, the $\Delta B$ in the winding mentioned in the literature is 0.2 T. Extrapolating the loss curves of the twisted samples in Fig. 5 to this $\Delta B$ values, one gets about 7.7 mW/Am. This indicates that further loss reducing mechanisms have still to be applied even with a square geometry, e.g. ceramic interfilamentary barriers, somewhat smaller conductor cross-sections, and smaller filament size. In both Figs 4 and 5, the losses of the samples with $l_t$ = 8 mm and $l_t$ = 4 mm are not very different. The reason might be that heavier twisting creates more bridging and gets lower effective resistivity, thus yielding higher coupling loses, according to eq. (1).

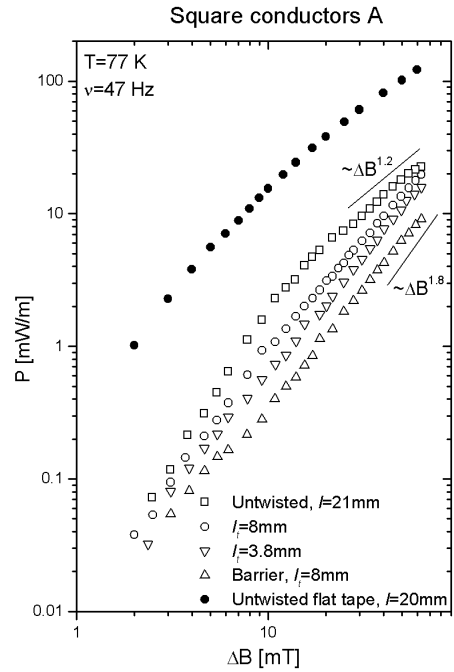

Fig. 4 Amplitude dependence at 47 Hz of power losses per meter of conductor length for wire A, the losses of a flat tape with aspect ratio16 in a magnetic field perpendicular to its wide side are displayed for comparison (solid points).

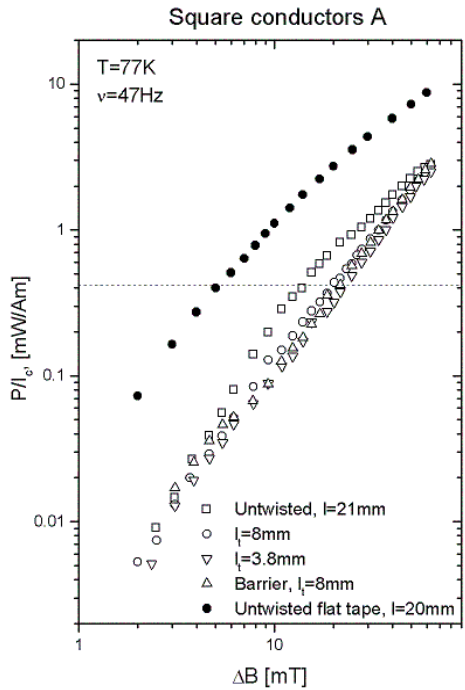

Fig. 5 Amplitude dependence at 47 Hz of power losses per ampere of critical current and meter of conductor length for wire A, the losses of a flat tape with aspect ratio16 in a magnetic field perpendicular to its wide side are displayed for comparison (solid points), the dashed line represents the tolerable loss level for transformer application.

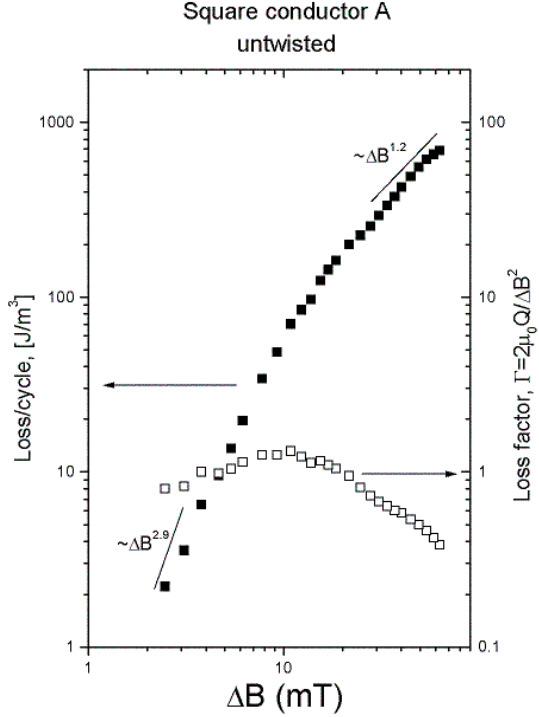

Fig. 6 Amplitude dependence at 47 Hz of loss per cycle and volume unit of untwisted conductor A (solid points) and loss factor $\Gamma$ (open points).

For a multifilamentary superconductor, it is also useful to know the composite penetration field $B_{ps}$ ($2B_{ps} = \Delta B_{ps}$) in the case of composite saturation. $\Delta B_{ps}$ can easily be calculated [1],[15]: for the untwisted square sample in Fig. 5 (showing saturation behavior), we have roughly calculated $\Delta B_{ps}$ assuming the filamentary zone as a square superconductor with an average $\overline{J_c} = I_c/F_z$, $F_z$ being the overall cross section of the filamentary zone, thus getting $\Delta B_{ps} \cong 9$ mT. Fig. 6 shows the measured loss per cycle and volume unit of the untwisted square wire with $l = 20$ mm. From this curve, the loss factor $\Gamma = 2\mu_0 Q/\Delta B^2$ (right scale in Fig. 6) is derived. The maximum of $\Gamma$ occurs at $\Delta B = 12$ mT, which is close to the calculated $\Delta B_{ps}$ value. We have also calculated the filament penetration field $\Delta B_{pf}$ for the central filament in one of the stacks of conductors A and B, assuming $\Delta B$ being perpendicular to the filament width and no composite saturation. For conductor A it is $\Delta B_{pf} \cong$ 5mT with $I_c = 8$ A, while for conductor B $\Delta B_{pf} \cong$ 5.3mT with $I_c = 13.5$ A. Despite the higher $I_c$ of conductor B, both values are very close, this is due to different values of stack aspect ratio and stack cross-section. Because of the arrangement of the filaments in these square conductors, the filament penetration field is larger than in a flat multifilamentary tape.

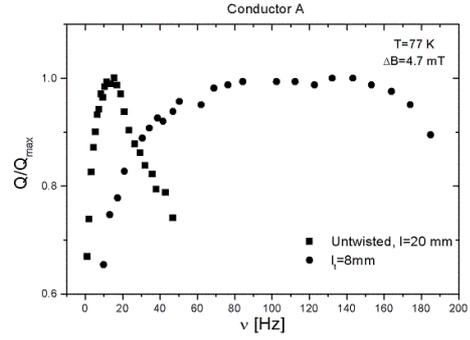

Fig. 7 Frequency dependence at $\Delta B = 4.7$ mT of the losses of conductor A. The losses of the untwisted wire with $l = 20$ mm have a maximum at 12.5 Hz, giving $\tau = 12.7$ ms. The losses of the wire with $l_t = 8$ mm have a maximum have a maximum at 135 Hz, giving $\tau = 1.2$ ms.

### 3.1.2 Coupling current decay time constant

Further, the coupling current decay time constant $\tau = 1/(2\pi\nu_m)$, with $\nu_m$ being the frequency where the losses are maximum, were measured with small $\Delta B < \Delta B_{ps}$. Fig 7 shows the loss curve for the untwisted square sample. From $\nu_m \cong 12.5$ Hz, one gets $\tau \cong 12.7$ ms. The decay time constant for an untwisted multifilamentary conductor of length $l$ is

$$\tau = \frac{1}{\pi^2}(1-N)\mu_0 \frac{l^2}{\rho_e} \qquad (18)$$

$N$ is the demagnetization factor of the composite, $N = \frac{1}{2}$ for round wire. For square wire one can also use approximately $N = \frac{1}{2}$. With the mentioned $\tau \cong$ 12.7ms and $l = 20$mm one gets for the effective resistivity $\rho_e \cong 2.1 \cdot 10^{-9}$ $\Omega$m compared to $\rho_{Ag} = 2.6 \cdot 10^{-9}$ $\Omega$m at 77K. Fig. 7 shows also the loss curve of the square sample with $l_t = 8$ mm. Here one find the maximum loss frequency at $\nu_m \cong 135$ Hz giving a $\tau$ of 1.2 ms. This value is comparable to the best $\tau$ values measured in twisted flat tapes with ceramic barriers and/or Ag-Au matrix [16],[10].

### 3.2 Square conductors B

### 3.1.2 Magnetic field losses at 47 Hz

In this section, we present the experimental loss results of the square multifilamentary conductor B, which differs from conductor A mainly by the twisting method, the superconductor filling factor and conductor dimensions of conductors A and B were similar but not fully identical (Table 1).

As both Figs 8 and 9 show, the losses in conductor B are still smaller than in conductor A. The untwisted square sample with $l = 12$ mm shows composite saturation, with losses being proportional to $\Delta B^n$ and $n \cong 1.0$ in the higher field range. For the twisted samples one finds approximately $n \cong 1.8$, indicating that saturation at 47Hz will only occur at still higher $\Delta B$ values. According to Figs 8 and 9, the extremely small twist length $l_t = 2$ mm does not further reduce the losses compared to $l_t = 4$ mm. This is probably due to filament bridging and has still to be investigated in detail. The sample with $l_t = 4$ mm has at 63 mT an overall power loss $P \cong 0.9$ mW/Am, being about 10 times smaller than the corresponding loss of the flat tape. The application loss limit $(P/I_c)_{tol}$ for a transformer is here fulfilled with $\Delta B \leq 43$ mT, even if this is a conductor with a pure Ag matrix.

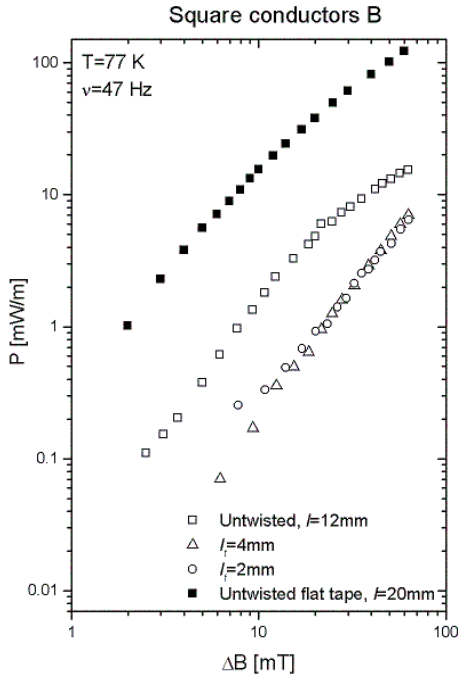

Fig. 8 Amplitude dependence at 47 Hz of power losses per meter of conductor length for wire B, the losses of a flat tape with aspect ratio16 in a magnetic field perpendicular to its wide side are displayed for comparison (solid points).

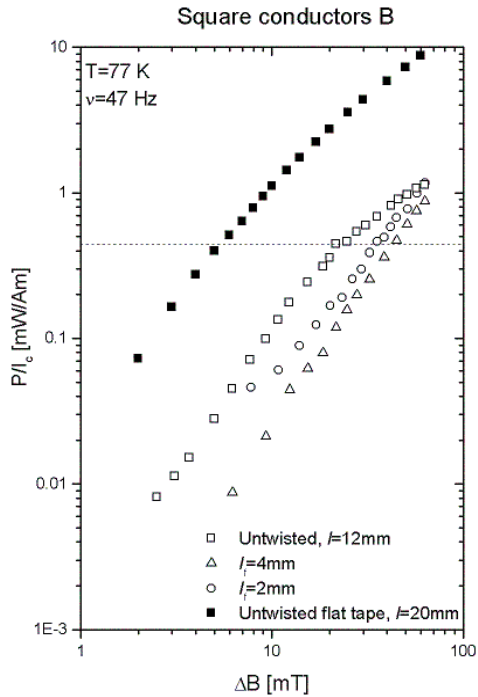

Fig. 9 Amplitude dependence at 47 Hz of power losses per ampere of critical current and meter of conductor length for wire B, the losses of a flat tape with aspect ratio16 in a magnetic field perpendicular to its wide side are displayed for comparison (solid points), the dashed line represents the tolerable loss level for transformer application

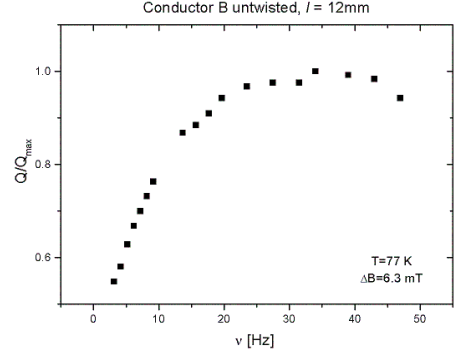

Fig. 10 Frequency dependence at $\Delta B = 6.3$ mT of the losses of untwisted conductor B with $l = 12$ mm, the losses have a maximum at 35 Hz, giving $\tau = 4.6$ ms.

The losses are even smaller than those of a flat tape with a Ag matrix having in addition interfilament ceramic barriers [16] in a perpendicular applied magnetic field, or those of a tape with a Ag-Au matrix having about 8 times larger $\rho$ than $\rho_{Ag}$. For a barrier conductor with $l_t = 13$ mm (B5 in [7]) we obtained 5.1 mW/Am at 48 Hz and $\Delta B = 63$ mT. Conductor B has thus about 6 times smaller losses. For comparisons, we had measured 3.1 mW/Am for a flat multifilamentary tape from BICC with a Ag-10wt%Au matrix, under the same conditions (unpublished). This value is about 4 times larger than in conductor B. Extrapolating in Fig. 9 the square conductor losses to $\Delta B = 0.2$ T, one roughly expect $P \cong 3$ mW/Am. This is still about 8 times larger than the transformer application loss limit. For $\Delta B = 0.1$ T, the losses are only about 5 times above the limit.

### 3.2.2 Coupling current decay time constant

The $\tau$ value of the untwisted sample of conductor B with $l = 12$ mm was also determined. Fig. 10 shows the measured frequency dependence of the loss per cycle at $\Delta B = 6.3$ mT. From $\nu_m \cong 35$ Hz, one get $\tau \cong 4.6$ ms, and using eq. (18), $\rho_e \cong 2 \cdot 10^{-9}$ $\Omega$m. From eq. (16) one could expect with this $\rho_e$ value $\tau \cong 0.5$ ms for the sample with $l_t = 4$ mm. On the other side the $l_t^2$-dependence of $\tau$ is not often not found in High-$T_c$ multifilamentary superconductors. In experiments with twisted flat tapes one gets typically $\tau \propto l_t^{1.6}$ assuming that $\rho_e$ is not affected by twisting. With this exponent, one would get for $l_t = 4$ mm $\tau \cong 0.9$ ms instead of the mentioned 0.5 ms.

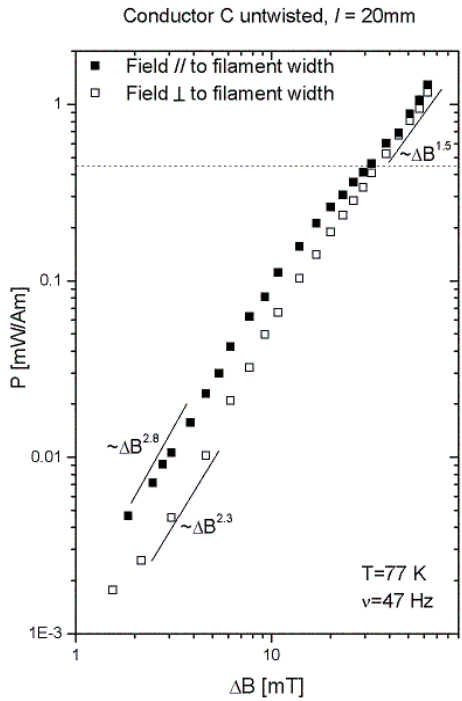

Fig. 11 Amplitude dependence at 47 Hz of power losses per ampere of critical current and meter of conductor length for wire C, the dashed line represents the tolerable loss level for transformer application.

### 3.3 Square conductor C

The square conductor C (Fig. 3) differs from the other square conductors A and B, as the filaments are not arranged in stacks being perpendicular to each other. However due to its simpler geometry, the losses can be analyzed more easily.

The losses at 47Hz were measured in two configurations, both for $\Delta B$ applied perpendicular and parallel to the wide side of the filaments. The results are displayed in Fig. 11. At the upper $\Delta B$ values, the losses in both directions are nearly identical, being proportional to $\Delta B^{1.3}$ for parallel $\Delta B$ and to $\Delta B^{1.5}$ for perpendicular $\Delta B$. These exponent values indicate that the conditions are close to full composite saturation. The absolute loss value in this conductor are also rather low, compared to those of the flat tapes. For $\Delta B \leq 34$ mT the losses are below the transformer loss limit of 0.42 mW/Am at 47 Hz. Also for this untwisted conductor with $l = 20$ mm, the frequency dependence of the loss per cycle at $\Delta B = 6.3$ mT was measured, and the $\tau$ values were determined (not shown). One gets for perpendicular $\Delta B$ $\tau \cong 4.25$ ms and for parallel $\Delta B$ $\tau \cong 4.3$ ms. This similar results may be due an interplay of somewhat different demagnetization factors of the filamentary zone and different $\rho_e$ values in both field directions. Due to the vicinity of $\nu_m \cong 35$ Hz to 47 Hz, There is in Fig. 11 in the low $\Delta B$ range a large $\Delta B^2$ contribution of the coupling currents to the total losses.

### 4 Conclusions

We have produced Bi,Pb(2223)/Ag multifilamentary prototype superconductors with square cross-section and a rectangular arrangement of filaments. As predicted by the theory, considerably reduced 50Hz magnetic field losses were measured in these square conductors compared to those of the flat Bi,Pb(2223)/Ag tapes in perpendicular magnetic field changes. This is due to the smaller composite width c in perpendicular applied $\Delta B$, as predicted by eq. (3). The losses of this wires are still higher by a factor of about 10 when compared to the one of the best tapes in a parallel field.

For applications where the magnetic field has a constant orientation and where it is possible to use tapes aligned in parallel to the magnetic field, tape have better performance both from the point of view of AC losses and transport properties, but for applications where a perpendicular component of the magnetic field cannot be avoided or applications in rotating field, the potential of square conductors is clearly demonstrated due to their smaller AC losses and the largely reduced angular anisotropy of the critical current due to the rectangular filament stacking [17].

By the application of further improvements to the square superconductors, the 50 Hz AC losses at the necessary $\Delta B$ value will thus be reduced below the tolerable loss level for High-$T_c$ transformers. The possible improvements are: introduction of highly resistive interfilamentary ceramic barriers in the matrix having at least the same quality as those applied so far in flat tapes, achievement of higher $I_c$ values, and furher reductions of overall conductor dimensions and of filament width.

The Critical current densities of the square wires analyzed in the present work were considerably lower than for flat Bi,Pb(2223) tapes, but the fundamental aspect concerning the influence of aspect ration on the AC losses are still valid. Further developments in our laboratory are under work to achieve square wires with higher $J_c$ values, the value of $J_c(77K, 0T) = 20$ kA/cm$^2$ being considered as achievable